\documentclass[11pt]{article}

\usepackage{amsmath}

\begin{document}

\newcommand{\N}{N\raise.7ex\hbox{\underline{$\circ $}}$\;$}

\thispagestyle{empty}

\begin{center}
{\bf

BELARUS NATIONAL ACADEMY OF SCIENCES

B.I. STEPANOV's  INSTITUTE OF PHYSICS

}
\end{center}

\vspace{30mm}

\begin{center}

{\bf  E.M. Ovsiyuk\footnote{or E.M. Bychkovskaya},   N.G.
Tokarevskaya,
 V.M. Red'kov\footnote{E-mail: redkov@dragon.bas-net.by} }

\end{center}

\vspace{5mm}

\begin{center}

{\bf SHAPIRO's PLANE WAVES IN SPACES OF CONSTANT CURVATURE  \\
AND SEPARATION OF VARIABLES IN REAL AND COMPLEX COORDINATES
 }

\end{center}


\thispagestyle{empty}

 \begin{quotation}

The aim of the article to  clarify   the status of Shapiro plane
wave solutions of the Schr\"{o}dinger's equation  in the frames of
the  well-known general method of separation of variables. To
solve  this task, we use the well-known cylindrical coordinates in
Riemann  and Lobachevsky spaces, naturally  related with Euler
angle-parameters. Conclusion  may be drawn: the  general  method
of separation of variables   embraces the  all plane wave
solutions; the plane waves in Lobachevsky  and Riemann space
consist of  a small part of the whole set of basis wave functions
of Schr\"{o}dinger equation.

In space of constant positive curvature  $S_{3}$, a complex analog
of horospherical coordinates of Lobachevsky space $H_{3}$ is
introduced. To parameterize  real space   $S_{3}$, two complex
coordinates  $(r,z)$ must obey additional restriction in the form
of  the  equation $r^{2} = e^{z-z^{*}} - e^{2z} $. The metrical
tensor of space $S_{3}$ is expressed in terms of  $(r,z)$  with
additional constraint, or through pairs  of conjugate variables  $
(r,r^{*})$ or  $(z,z^{*})$; correspondingly exist three different
representations for Schr\"{o}dinger Hamilto\-nian.
 Shapiro plane waves are determined and explored as solutions of  Schr\"{o}dinger
 equation in complex horisperical coordinates of $S_{3}$.
 In particular, two oppositely directed plane waves may be presented
 as exponentials in conjugated  coordinates.
       $\Psi_{-}= e^{-\alpha z}$ and  $\Psi_{+}= e^{-\alpha z^{*}}$.
Solutions constructed
 are single-valued, finite, and continuous functions in spherical space and correspond to discrete energy levels.

\end{quotation}

\vspace{5mm}

Published in:

\vspace{5mm}

 E.M.  Bychkovskaya,  N.G. Tokarevskaya,  V.M. Red'kov.
  Shapiro's plane waves in spaces of constant
curvature  and separation of variables in real and complex
coordinates. Nonlinear Phenomena in Complex Systems. {\bf  12}, 1--15  (2009).

\newpage

\subsection*{1. Introduction }

It is well known that in field theory of elementary particles the
most used elementary solutions are  plane  waves. However, in any
curved  space-time such simple plane wave solutions do not exist.
Instead, only very symmetrical space-time models may have some
analogues of such waves. In particular, there are known Shapiro's
plane waves \cite{Shapiro-1962}, in the Lobachevsky space $H_{3}$,
they were introduced in the context of expansion of scattering
amplitudes in  relativistic spherical functions. In the context of
quantum mechanics  for Schr\"{o}dinger  particle these wave
functions are eigenfunctions of projection of generator of
displacements on Lobachevsky space on arbitrary direction
described by 3-vector (about the notation see below):
\begin{eqnarray}
 \Psi  =  e^{-i\epsilon t} \;
(u_{0} + {\bf n} \; {\bf u})^{\alpha}  \; , \qquad  ({\bf P} {\bf
n}) \; \Psi = \alpha \; \Psi \; , \qquad \alpha = 1 \; \mp \; i \;
\sqrt{2\epsilon -1} \;. \label{1}
\end{eqnarray}

Solutions of that type can be constructed explicitly in arbitrary
coordinate system of the space $H_{3}$ In
\cite{Bychkovskaya-2006},  the plane Shapiro waves were considered
on the base of  generalized cylindrical coordinates. Recently,
that extended plane waves  were used in  solving  the scattering
problem for Schr\"{o}dinger particle on Coulomb center in
Lobachevsky space \cite{Bogush-Kurochkin-Otchik-2003}. Besides,
solutions of that type were constructed for Maxwell equations in
Lobachevsky space \cite{Bychkovskaya-2006},
\cite{Bogush-Kurochkin-Otchik-Bychkovskaya-2006}, they were
specified in two coordinate systems:  cylindric and horospheric.
In should be noted that the
 scalar wave equation in these coordinates in Lobachevsky model was extensively investigated many years ago
 by Vilenkin and Smorodinsky \cite{Vilenkin-Smorodinsky-1964}.

 The status
of Shapiro plane waves for both models $H_{3}$ and $S_{3}$
  from the  viewpoint of the well-known general method of separation of variables was explored on the  base of
  corresponding cylindrical coordinates.
  Conclusion  was been drawn:
the  general  method of separation of variables   embraces the
all plane wave solutions; the plane waves in Lobachevsky  and
Riemann space  consist of  a small part of the whole set of basis
wave functions of Schr\"{o}dinger equation.

In contrast to cylindric systems for $H_{3}$ and $S_{3}$, the
orispherical  system exists only in Lobachevsky space, the  full
list of appropriate coordinate system relevant to method of
separation of variables  was given by Olevsky \cite{Olevsky-1950}:
 in the model $H_{3}$ we have 34 systems, in
the model $S_{3}$ we have  only 6 ones. However, in
\cite{Bogush-Otchik-Red'kov-2006}, it was pointed out one method
to have some analogues for all 34 coordinate systems of the space
$H_{3}$ in the space model $S_{3}$ as well: it suffices to permit
the use of complex coordinates  in real space $S_{3}$.
 The aim of the present paper is to develop this approach for  complex orispherical coordinates
 in real spherical space $S_{3}$ on this base to examine the plane waves Shapiro's type
 in spherical model.

It should be stressed that special interest may have the task of
extending the general method of separation of variables to complex
curvilinear coordinates.

\subsection*{2. Shapiro's plane waves in Lobachevsky space and cylindric coordinates
}

In Olevsky paper  \cite{Olevsky-1950}, among  34 coordinate systems  we see the  following one
\begin{eqnarray}
dS^{2} =  dt^{2} -  dr^{2}  - \mbox{sh}^{2} r\; d\phi^{2} -
\mbox{ch}^{2} r\;  d z^{2} \; , \nonumber
\\
r \in [0 , + \infty )\; , \qquad z \in (-\infty , + \infty )\; ,
\phi \in [ 0 , 2\pi ]\; , \nonumber
\\
u_{1} = \mbox{sh}\; r \; \cos \phi \; , \; u_{2} = \mbox{sh}\;
r \; \sin \phi \; , \;
u_{3} = \mbox{ch}\;r\; \mbox{sh}\; z \; ,  \;  u_{0} =
\mbox{ch}\;r\; \mbox{ch}\; z \; . \label{2.1}
\end{eqnarray}

\noindent all coordinates are dimensionless. In the limit of vanishing curvature, relations (\ref{2.1}) will define
 the ordinary cylindric coordinates.

In quasi-Cartesian coordinates $u_{a}$ of $H_{3}$-model, Shapiro's  solutions are determined by
\begin{eqnarray}
u_{0}^{2} - {\bf u}^{2} =1 \; , \qquad \Psi =  e^{-i\epsilon t} \;
(u_{0} + {\bf n} \; {\bf u})^{\alpha}  \; , \label{2.2}
\end{eqnarray}

\noindent where ${\bf n}$ stands for any unit 3-vector.
Taking orientation vector  ${\bf n}$ along third axis we get
\begin{eqnarray}
{\bf n} = (0,0, +1)\; , \qquad \Psi =  e^{-iE t/\hbar } \; (u_{0}
+ u_{3})^{\alpha} =
  e^{-iE t/\hbar} \; (\;
\mbox{ch} \; r \; e^{z} \; ) ^{\alpha} \; . \label{2.3}
\end{eqnarray}

\noindent It is the matter of simple calculation to show that eq. (\ref{2.3}) defines
an exact solution of Schr\"{o}dinger equation in coordinates (\ref{2.1}):
\begin{eqnarray}
  i  \hbar  \;  \partial  _{t }     \Psi = H \Psi \; ,
\qquad H = {1 \over 2M\rho^{2}}\; \left [ \; {i \hbar  \over
\sqrt{-g} } \;
\partial_{k} \sqrt{-g}  \; (- g^{kl}(x)
)\;i\; \hbar \; \partial  _{l }  \; \right ]= \nonumber
\\
=- { \hbar^{2} \over 2M\rho^{2} } \left [ \; {1 \over \mbox{sh}\;
r \; \mbox{ch}\; r }\;  {\partial \over \partial r} \mbox{sh}\; r
\; \mbox{ch}\; r  {\partial \over \partial r} + {1 \over
\mbox{sh}^{2} r } {\partial^{2} \over \partial \phi^{2}} + {1
\over \mbox{ch}^{2} r } {\partial^{2} \over \partial z^{2}}  \;
\right ] \; ;
 \label{2.4}
 \end{eqnarray}

\noindent indeed, eq. (\ref{2.4}() reads
\begin{eqnarray}
 \; {1 \over \mbox{sh}\; r \; \mbox{ch}\; r }\;  {\partial \over \partial r}
\mbox{sh}\; r \; \mbox{ch}\; r  {\partial \over \partial r}
\mbox{ch} ^{\alpha} r  + \alpha^{2}  \;  \mbox{ch} ^{\alpha-2} r +
2\epsilon\;    \mbox{ch}^{\alpha}  r = 0 \; , \qquad \epsilon = {E
\over \hbar^{2} / M \rho^{2}}  \;  ;  \nonumber
\end{eqnarray}

\noindent which becomes identity when parameter $\alpha$  obeys
$\alpha^{2} +2\alpha + 2\epsilon = 0 \;  .
$
 Thus, Schr\"{o}dinger equation has the following solutions
\begin{eqnarray}
\Psi = e^{-iE t/\hbar} \;  \mbox{ch}^{\alpha} r \; e^{ \alpha z}
\; , \qquad
 \alpha =\alpha _{\pm}=  -1 \pm i\; \sqrt{2\epsilon -1}  \;  .
\label{2.6}
\end{eqnarray}

\noindent where $(2\epsilon -1 )\geq 0$ because in space $H_{3}$
the energy spectrum of a free particle starts with a definite
minimal value  $
 \epsilon > {1 \over 2} \; , \; \mbox{or} \;  E > \hbar^{2} / 2M\rho^{2} \;$.
\noindent The limiting procedure to flat space is realized according to
\begin{eqnarray}
e^{\alpha Z} = \mbox{exp} \; [  \; {1 \over \rho}  ( -1 \pm \sqrt{1- {
2E \;  M \rho^{2}\over \hbar^{2} } } \; ) \; \rho z \; ] \; ,
\qquad \mbox{exp} \;  [\; \pm i \sqrt{ 2EM \over \hbar^{2}}\; z ]
= e^{\pm iPz /\hbar} \;  , \nonumber
\end{eqnarray}

\noindent Asymptotic  behavior of the solution
is given by

\vspace{3mm}
${\bf n} = (0,0,+1)\; , $
\begin{eqnarray}
(\mbox{ch} ^{\alpha_{\pm}} r ) _{r \rightarrow 0} \longrightarrow 1 \; , \qquad
(\mbox{ch} ^{\alpha_{\pm}} r ) _{r \rightarrow + \infty } \longrightarrow  \infty  \; ,
\nonumber
\\
z \rightarrow + \infty \; , \; e^{\alpha_{\pm}z} \rightarrow 0
\; ,  \qquad
z \rightarrow - \infty \; , \; e^{\alpha_{\pm}z} \rightarrow
\infty  \; . \label{2.8b}
\end{eqnarray}

\noindent When starting with opposite orientation
\begin{eqnarray}
{\bf n} = (0,0,-1)\; , \qquad \qquad \Psi =  e^{-iE t/\hbar } \; (u_{0} -
u_{3})^{\alpha} = \nonumber
\\
= e^{-iE t/\hbar} \; (\mbox{ch}\;r\; \mbox{ch}\; z -
\mbox{ch}\;r\; \mbox{sh}\; z )^{\alpha} =  e^{-iE t/\hbar} \;
\mbox{ch} ^{\alpha} \; r \; e^{- \alpha z} \; , \label{2.9a}
\end{eqnarray}

\noindent we  arrive at the other asymptotic behavior in $z$-variable:
\begin{eqnarray}
z \rightarrow + \infty \; , \;\;  e^{-\alpha_{\pm}z} \rightarrow
\infty  \;   \; , \qquad
z \rightarrow - \infty \; , \;\;  e^{-\alpha_{\pm}z} \rightarrow
0  \; .  \label{2.9b}
\end{eqnarray}

It may be easily shown that Shapiro  plane waves satisfy the eigenvalue equation:
\begin{eqnarray}
({\bf P} \; {\bf n} ) \; \Psi = \alpha \; \Psi \; , \qquad \Psi =
e^{-i\epsilon t} \; (u_{0} + {\bf n} \; {\bf u})^{\alpha} \; ,
\label{2.10a}
\end{eqnarray}

\noindent where ${\bf P}$ plays a role of operator  momentum in $H_{3}$.
In accordance with  symmetry of spaces of constant curvature, the operators
 $\vec{P}, \;
\vec{L}$ obeys Lie algebra of  $SO(3.1)$, respectively:
and $SO(4)$:
\begin{eqnarray}
{\bf P} = -i (1 \mp {\bf q} \bullet {\bf q}) \; {\partial \over
 \partial {\bf q}} \;\; , \;\ \vec{L} = [\; {\bf q} \;{\bf P}\;] = -i [ {\bf q} , \;{\partial \over {\bf q}}  ] \; ,
\label{A}
\end{eqnarray}

\noindent  the sign $-$ corresponds to  $H_{3}$-model, the sign $+$ is referred to  $S_{3}$-model;
vector variable is defined by
\begin{eqnarray}
{\bf q} = {   {\bf u} \over u_{0} } \;, \qquad u_{0}^{2} - {\bf
u}^{2} =1 \; ; \nonumber
\end{eqnarray}

\noindent the quantities  $\vec{L}$ and $\vec{P}$ are measured in  $\hbar$ and
$\hbar / R$; $\bullet$ respectively. To  verify eq. (\ref{2.10a}), let us translate the above solutions to
${\bf q}$-variables:
\begin{eqnarray}
u_{0} = {1 \over + \sqrt{1 - q^{2}} }  \; , \qquad {\bf u} = {
{\bf q} \over  + \sqrt{1 -q^{2}}} \; , \nonumber
\\
\Psi =  e^{-i\epsilon t} \; (1 - q^{2} )^{-\alpha /2}   \; (1 +
{\bf n} \; {\bf q})^{\alpha} \; . \label{2.10b}
\end{eqnarray}

\noindent Taking into account relation
\begin{eqnarray}
({\bf P} \; {\bf n} ) \; \Psi  =-i ( n_{i} { \partial \over
\partial q_{i}}   -
 n_{j}q_{j} \;q_{i} {\partial \over \partial q_{i}} ) \;\left [
  (1 - q^{2} )^{-\alpha /2}   \; (1 + {\bf n} \; {\bf q})^{\alpha} \right ]  =
\nonumber
\\
= -i \; (1 - q^{2} )^{-\alpha /2}   \; (1 + {\bf n} \; {\bf
q})^{\alpha} \; [\; \alpha \; { (n_{i} q_{i})  \over  1 - q^{2} }
\;  + { \alpha   \over 1 + {\bf n} \; {\bf q} } -
 {(n_{j} q_{j})\;
\alpha q^{2} \over  1 - q^{2} } \;  - { \alpha  \; (n_{j} q_{j})
\;
 (q_{i}n_{i})     \over 1 + {\bf n} \; {\bf q} } \; ]\; ,
\nonumber
\end{eqnarray}

\noindent after simple calculation we  arrive at
\begin{eqnarray}
( {\bf P} \; {\bf n} )  \Psi = -i\alpha \; \Psi \;   . \label{2.10c}
\end{eqnarray}

\noindent Relating the wave function with Schr\"{o}dinger solution,
we can establish connection between the eigenvalue  $(-i\alpha)$ and the energy value
 $\epsilon$ (see (\ref{2.6})):
\begin{eqnarray}
-i  \alpha =-i \alpha _{\pm}=  +i \pm  \sqrt{2\epsilon -1}
 \; ,
\label{2.10e}
\end{eqnarray}

\noindent in usual units this relationships exhibits correct behavior when vanishing the curvature:
\begin{eqnarray}
-i\alpha \; {\hbar \over \rho} =  \; +i {\hbar \over \rho}  \pm  \sqrt{2EM - {\hbar^{2} \over \rho^{2}} } \;\;
 \qquad
 \longrightarrow
 \qquad
  \pm \sqrt{2EM} \; \;\; \mbox{при} \;\;  \rho \rightarrow \infty \; .
\nonumber
\end{eqnarray}

 Taking the vector  ${\bf n}$ in three different ways, one can construct eigenfunctions of
 $P_{1}, P_{2},P_{3}$ respectively:
\begin{eqnarray}
{\bf n}=(1,0,0)\; , \qquad
 \hat{P}_{1}   \Psi_{1} = -i \alpha \; \Psi_{1} \; ,
\qquad \Psi_{1}  =  e^{-i\epsilon t}  (1 - q^{2} )^{-\alpha /2} (1
+  q_{1})^{\alpha} \; . \nonumber
\\
{\bf n}=(0,1,0)\; , \qquad
 \hat{P}_{2}   \Psi_{1} = -i \alpha \; \Psi_{2} \; ,
\qquad \Psi_{2}  =  e^{-i\epsilon t}  (1 - q^{2} )^{-\alpha /2} (1
+  q_{2})^{\alpha} \; . \nonumber
\\
{\bf n}=(0,0,1)\; , \qquad
 \hat{P}_{3}   \Psi_{1} = -i \alpha \; \Psi_{3} \; ,
\qquad \Psi_{3}  =  e^{-i\epsilon t}  (1 - q^{2} )^{-\alpha /2} (1
+  q_{3})^{\alpha} \; . \label{2.11}
\end{eqnarray}

\noindent For flat space model, one can multiply three elementary functions
\begin{eqnarray}
\Psi_{1}^{0} = e^{ik_{1}x}\; , \qquad \Psi_{2}^{0} = e^{ik_{2}y}\;
, \qquad \Psi_{3}^{0} = e^{ik_{3}z}\; , \nonumber
\end{eqnarray}

\noindent and produce an eigenfunction of the vector-operator
$\hat{{\bf P}}$. In the curved space, such method does nod work because of no-commutativity
components of the momentum -- see  (\ref{A}).

Let us find expression for  $P_{3}$ in coordinates  $x^{i}=
(r,\phi,z)$. Starting from
\begin{eqnarray}
q_{1} = {\mbox{th}\; r \over \mbox{ch}\; z}  \; \cos \phi \; ,
\qquad q_{2} = {\mbox{th}\; r \over \mbox{ch}\; z} \; \sin \phi \;
, \qquad q_{3} = \mbox{th}\; z \; ; \nonumber
\\
\mbox{th}\; z =  q_{3}\; ,  \qquad  \mbox{ch}\; z = {1 \over
\sqrt{ 1 - q_{3}^{2}  } } \; , \qquad \mbox{th}\; r =  \sqrt{
{q_{1}^{2} + q_{2}^{2} \over 1 - q_{3}^{2}}} \; ,\nonumber
\\
\cos \phi = { q_{1} \over \sqrt{ q_{1}^{2} + q_{2}^{2} } } \; ,
\qquad \sin \phi = {q_{2} \over \sqrt{ q_{1}^{2} + q_{2}^{2}  }}
\; . \label{2.12}
\end{eqnarray}

\noindent for   $P_{3}$  we get
\begin{eqnarray}
P_{3} =
 -i \; [\; {\partial  x^{i} \over  \partial q_{3}} {\partial \over \partial x^{i}} -
  q_{3}\;
( q_{1}  {\partial x^{i} \over \partial q_{1}}  + q_{2}  {\partial
x^{i} \over \partial q_{2}}  + q_{3}  {\partial x^{i} \over
\partial q_{3}}  ) {\partial \over  \partial x^{i}}  \; ] \; .
\label{2.13a}
\end{eqnarray}

\noindent Taking into account identities
\begin{eqnarray}
{\partial r \over \partial q_{1}} = \mbox{ch}^{2} r \; \mbox{ch}\;
z \; \cos \phi\; , \qquad {\partial r \over \partial q_{2}} =
\mbox{ch}^{2}r \; \mbox{ch}\; z \; \sin \phi\; , \qquad {\partial
r \over \partial q_{3}}= \mbox{sh}\; r\; \mbox{ch}\;r\;
\mbox{sh}\; z\; \mbox{ch}\;z\; , \nonumber
\\
{\partial \phi \over \partial q_{1}} = - { \mbox{ch}\; z \over
\mbox{th}\; r } \; \sin \phi\; , \qquad {\partial \phi \over
\partial q_{2}} = + { \mbox{ch}\; z \over \mbox{th}\; r } \; \cos
\phi\; , \qquad {\partial \phi \over \partial q_{3}} = 0 \; ,
\nonumber
\\
{\partial z \over \partial q_{1}} = 0\; , \qquad {\partial z \over
\partial q_{2}} = 0\; , \qquad {\partial z \over \partial q_{3}} =
\mbox{ch}^{2} z \; . \nonumber
\end{eqnarray}

\noindent after simple calculation we arrive at the formula
\begin{eqnarray}
P_{3} = -i {\partial \over  \partial z} \; . \label{2.13e}
\end{eqnarray}

\noindent Immediately, one produces the identity expected:
\begin{eqnarray}
-i {\partial \over  \partial z}\; [\;  e^{-iE t/\hbar} \;
\mbox{ch}^{\alpha} \; r \; e^{\alpha z}  \; ]= -i \alpha  \;[\;
 e^{-iE t/\hbar} \;  \mbox{ch}^{\alpha} \; r \; e^{\alpha z}  \; ] \; .
 \nonumber
 \end{eqnarray}

Now, let us relate the above plane wave solutions  (\ref{2.6})  and (\ref{2.9a}) with
all solutions of Schr\"{o}dinger equation constricted within general method
of separation of variables in cylindric coordinates:
\begin{eqnarray}
\Psi = e^{-iEt /\hbar} \;  e^{im\phi}  \;  e^{\alpha z} \; G(r) \;
, \nonumber
\\
\; [\; {d^{2} \over dR^{2}} + (  {\mbox{ch}\; r \over \mbox{sh}\;
r} +  {\mbox{sh}\; r \over \mbox{ch}\; r}) \; {d \over dR}
 -
{m^{2} \over \mbox{sh}^{2} r }  + {\alpha^{2} \over \mbox{ch}^{2}
r }  \; + 2 \epsilon \;  ]\;  G(r)   = 0 \; . \label{2.15}
\end{eqnarray}

\noindent In  new variable
$\mbox{ch}^{2}r = y, \; y \in
[1 , + \infty )$
eq.  (\ref{2.15})  takes the form
\begin{eqnarray}
[\; 4y(y-1) {d^{2} \over dy^{2}} + 4 (2y-1) {d \over dy} -{m^{2}
\over y-1} + {\alpha^{2} \over y} +2\epsilon \; ] \; G = 0 \; .
\label{2.16}
\end{eqnarray}

\noindent With the use of the following substitution
\begin{eqnarray}
G(y) =  (y-1)^{a}\; y^{b} \; Y (y) \; . \nonumber
\end{eqnarray}

\noindent we get
\begin{eqnarray}
4(y-1) y  \; Y''  + 4 [ \; 2(a+b+1)y - (2b+1)]\; Y' + \nonumber
\\
+ \{ \;
 [ 4a(a-1)  + 8ab  + 4b(b-1) + 4a +4b  +4 a + 4b +2\epsilon ] \; +
\nonumber
\\
+ [ -4b(b-1)  -4b  + \alpha^{2}   ] \;{1 \over y} +  [  4a(a-1)
+4a  -m^{2} ] \; {1 \over y-1 } \;  \}\; Y = 0 \; . \label{2.20}
\end{eqnarray}

\noindent Let us require
\begin{eqnarray}
-4b(b-1)  -4b  + \alpha^{2}  = 0 \; , \qquad b = \pm {\alpha \over
2} \; ; \nonumber
\\
 4a(a-1) +4a  -m^{2} = 0 \; , \qquad a =  \pm {m \over 2} \; .
\label{2.21}
\end{eqnarray}

\noindent then eq. (\ref{2.20})  reduces to
\begin{eqnarray}
4(y-1) y  \; Y''  + 4 [ \; 2(a+b+1)y - (2b+1)]\; Y' + \nonumber
\\
+  [ 4a(a+1)  + 8ab  + 4b(b+1)  + 2 \epsilon ] \;  Y = 0 \; ;
\label{2.22}
\end{eqnarray}

\noindent that is of  the hypergeometric type
\begin{eqnarray}
(y-1) y  \; Y''  +  [ \;  (A + B + 1)y  - C  \; ]\; Y' +  AB \;  Y
= 0 \; , \nonumber
\end{eqnarray}

\noindent if
\begin{eqnarray}
C = 2b+1 \; , \qquad  A + B =  2a+2b+1 \; , \qquad
 A B =  a (a+1)  + 2 ab  + b (b+1)  +  \epsilon/2 \; ,
\nonumber
\end{eqnarray}

\noindent or
\begin{eqnarray}
AB = (a+b+1/2)^{2} - {(1-2\epsilon) \over 4} \;, \qquad A + B =
2(a+b+ 1/2) \; . \nonumber
\end{eqnarray}

Thus, the problem of wave functions is solved:
\begin{eqnarray}
G(y) =  (y-1)^{a}\; y^{b} \; F(A , B, C ; y )  \; , \qquad y =
\mbox{ch}^{2} r \; , \nonumber
\\
 a =\pm {\mid m \mid \over 2} \; , \qquad b = \pm {\alpha \over 2} \; , \qquad  C = 2b +1 \; ,
\nonumber
\\
\; A = a+b+ {1 \over 2}   \pm i \; {\sqrt{ 2 \epsilon -1} \over 2}
\; , \qquad B = a+b+ {1 \over 2}   \mp  i\; {\sqrt{ 2 \epsilon-1}
\over 2} \; . \label{2.24}
\end{eqnarray}

Let us restrict ourselves and consider a part of solutions (\ref{2.24}):
\begin{eqnarray}
a =\pm  m / 2 = 0 \; , \qquad  b = + \alpha /2
\nonumber
\\
\Psi (r,z) =  e^{\alpha z} \;  (\mbox{ch}\; r) ^{\alpha}  \; F(A ,
B, C ; \mbox{ch}^{2}r )  \; , \nonumber
\\
 2A = \alpha + 1    \pm i \; \sqrt{ 2 \epsilon -1}  \; , \qquad
 2B = \alpha + 1   \mp  i\;  \sqrt{ 2 \epsilon-1}  \; .
\label{2.29a}
\end{eqnarray}

\noindent
Requiring $A=0$ or   $B=0$, we arrive at
\begin{eqnarray}
\underline{A=0,} \qquad  \alpha  = - 1    \mp  i \; \sqrt{ 2
\epsilon -1} \; , \nonumber
\\
F(0 , B, C ; \mbox{ch}^{2}r ) =1 \; , \qquad \Psi (r,z) =
e^{\alpha z} \;  (\mbox{ch}\; r) ^{\alpha}    \; ; \nonumber
\end{eqnarray}
\begin{eqnarray}
\underline{B=0,}  \qquad   \alpha  = - 1    \mp  i \; \sqrt{ 2
\epsilon -1} \; , \nonumber
\\
 F(A , 0, C ; \mbox{ch}^{2}r ) =1\; ,  \qquad
 \Psi (r,z) =  e^{\alpha z} \;  (\mbox{ch}\; r) ^{\alpha}  \; ;
\label{2.29b}
\end{eqnarray}

\noindent which coincides with (\ref{2.6}).

There exists another symmetric  variant
\begin{eqnarray}
a =\pm  m / 2 = 0 \; , \qquad   2b = - \alpha :
\nonumber
\\
 \Psi (r,z) =  e^{\alpha z} \;
(\mbox{ch}\; r) ^{-\alpha}  \; F(A , B, C ; \mbox{ch}^{2}r )  \; ,
\nonumber
\\
 2A = -\alpha + 1    \pm i \; \sqrt{ 2 \epsilon -1}  \; , \qquad
 2B = -\alpha + 1   \mp  i\;  \sqrt{ 2 \epsilon-1}  \; .
\label{2.30a}
\end{eqnarray}

\noindent which leads to
\begin{eqnarray}
\underline{A=0,} \qquad  \qquad \alpha  = + 1    \mp  i \; \sqrt{
2 \epsilon -1} \; ,
 \qquad
\Psi (r,z) =  e^{\alpha z} \;  (\mbox{ch}\; r) ^{-\alpha}    \; ,
\nonumber
\\
\underline{B=0,}  \qquad \qquad   \alpha  = + 1    \mp  i \;
\sqrt{ 2 \epsilon -1} \; , \qquad  \Psi (r,z) =  e^{\alpha z} \;
(\mbox{ch}\; r) ^{-\alpha}  \; ; \label{2.30b}
\end{eqnarray}

\noindent this solution coincides with (\ref{2.9a}).

Conclusion  may be drawn:
the  general  method of separation of variables   embraces the  all plane wave solutions;
the plane waves in Lobachevsky space  consist of  a small part
of the whole set of basis wave functions of Schr\"{o}dinger equation.

\subsection*{3.  Plane wave in spherical space and separation of variables
}

Let us extend results of previous section  to the case of space of positive constant curvature, spherical space
$S_{3}$. In  this space there exist \cite{Olevsky-1950} similar cylindric coordinates
\begin{eqnarray}
dl^{2} = [\; d \rho ^{2}  + \sin ^{2} \rho  d \phi ^{2} + \cos
^{2} \rho  dz^{2} \; ] \; , \;\; \rho  \in  [ 0, \pi /2 ]\; , \;\;
\phi , z \in  [ - \pi  ,  - \pi  ] \; ; \nonumber
\\
u_{0} = \cos  \rho  \cos  z  \; , \;\;  u_{3} = \cos  \rho  \sin z
\; , \;\; u_{1} = \sin  \rho  \cos  \phi  \; , \;\; u_{2} = \sin
\rho  \sin  \phi \; . \label{3.1}
\end{eqnarray}

\noindent Shapiro's wave in this space model is given by
\begin{eqnarray}
u_{0}^{2} + {\bf u}^{2} =1 \; , \qquad \Psi =  e^{-i E t / \hbar }
\; (u_{0} + i\; {\bf n} \; {\bf u})^{\alpha}  \; . \label{3.2}
\end{eqnarray}

\noindent
Taking the vector  ${\bf n}$  along third axis, we get
\begin{eqnarray}
{\bf n} = (0,0, +1)\; , \qquad \Psi =  e^{-iE t / \hbar  } \;
(u_{0} + i u_{3})^{\alpha} =    e^{-iE t/\hbar} \;  \cos ^{\alpha} \rho \;
e^{i \alpha z} \;  . \label{3.3}
\end{eqnarray}

\noindent This function satisfies Schr\"{o}dinger equation in $S_{3}$:
\begin{eqnarray}
 \; {1 \over \sin \rho \cos \rho }   {\partial \over \partial \rho}
\sin \rho \cos\rho   {\partial \over \partial \rho } \cos^{\alpha}
\rho  - \alpha^{2}   \cos^{\alpha-2} \rho  + 2\epsilon
\cos^{\alpha}  \rho  = 0 \; , \nonumber
\end{eqnarray}

\noindent which is an identity  if
\begin{eqnarray}
\alpha^{2} +2\alpha - 2\epsilon = 0 \; , \qquad  \alpha =\alpha _{\pm}=  -1 \pm  \sqrt{2\epsilon + 1}
 \; .
\label{3.5a}
\end{eqnarray}

\noindent Thus, the plane wave solutions to Scr\"{o}dinger equation are
\begin{eqnarray}
\Psi^{\alpha_{\pm}}_{+} = e^{-iE t/\hbar} \;  \cos^{\alpha} \rho
\; e^{ i \alpha z} \; , \qquad \alpha _{\pm}=  -1 \pm
\sqrt{2\epsilon +1}
 \;  .
\label{3.5b}
\end{eqnarray}

\noindent Let us require the wave functions to be finite, continuous and single-valued function of point in space $S_{3}$.
From demand of periodicity in variable $z$  we get
\begin{eqnarray}
 \sqrt{2\epsilon +1} = n =   + 1, + 2, ... \;\; \Longrightarrow \;\; \epsilon = {n ^{2}-1 \over 2} \; ,
\nonumber
\\
\alpha_{-} = -1 -n  = -2, -3, ... \; , \qquad \alpha_{+} = -1 + n
= 0 , +1, +2, ...
 \label{3.6a}
\end{eqnarray}

\noindent One should take special attention to the value
 $\alpha_{+}=0$, when we obtain a very special wave function at $\epsilon = 0$:
 \begin{eqnarray}
\Psi^{\alpha=0} _{+} = (\cos \rho e^{+ i z} )^{0} = 1 \; ,
 \label{3.6b}
\end{eqnarray}

\noindent which may be associated with uniform  distribution of the particle over the whole space
 $S_{3}$. To test the functions constructed we must recall peculiarities of parametrization of $S_{3}$
 by cylindric coordinates:
The first peculiarity is
\begin{eqnarray}
\rho = 0 \; , \qquad  (0, \phi, z) \;\;  \Longleftrightarrow \;\;
\cos^{2}z + \sin ^{2} z = 1 \; ; \nonumber
\end{eqnarray}

\noindent on  this closed  the function (\ref{3.5b}) behaves correctly
\begin{eqnarray}
\Psi (\rho \rightarrow +0, \phi, z)  = e^{-iE t/\hbar} \;
 (\cos \rho \rightarrow +1)^{\alpha_{+}}  e^{ i \alpha_{+} z} =
e^{ i \alpha z_{+}} \; , \qquad
 \alpha_{+}  \in  + 1, + 2, ...
\nonumber
\\
\Psi (\rho \rightarrow +0, \phi, z)  = e^{-iE t/\hbar} \;  (\cos
\rho \rightarrow +1)^{\alpha_{-}}  e^{ i \alpha_{-} z} = e^{ i
\alpha z_{-}} \; , \qquad
 \alpha_{-} \in  - 1, - 2, ...
\nonumber
\end{eqnarray}

\noindent The second peculiarity is
\begin{eqnarray}
\rho = \pi /2 \; , \qquad (\pi/2, \phi, z)  \;\;
\Longleftrightarrow \;\; \cos^{2} \phi + \sin^{2} \phi = 1 \; .
\nonumber
\end{eqnarray}

\noindent the solutions constructed behave themselves as follows
\begin{eqnarray}
\Psi (\rho \rightarrow +\pi/2 , \phi, z)  = e^{-iE t/\hbar} \;
 (\cos \rho \rightarrow +0)^{\alpha_{+}}  e^{ i \alpha_{+} z} =
0  \;  , \qquad
 \alpha_{+} \in  + 1, + 2, ...
\nonumber
\\
\Psi (\rho \rightarrow + \pi/2 , \phi, z)  = e^{-iE t/\hbar} \;
 (\cos \rho \rightarrow +0)^{\alpha_{-}}  e^{ i \alpha_{-} z} =
\infty \;  e^{ i \alpha_{-} z}  \; , \qquad
 \alpha _{-}\in  - 1, - 2, ...
\nonumber
\end{eqnarray}

Therefore, all solutions with negative  $\alpha$ must be rejected as being discontinuous on  the whole
closed line on the sphere $S_{3}$ specified  by  $\rho = \pi/2$.
Thus, physical solutions to Schr\"{o}dinger equations on the sphere
are
\begin{eqnarray}
\Psi_{+}^{\alpha_{+}} =( \cos \rho e^{+iz})^{\alpha_{+}} \; .
\label{3.6c}
\end{eqnarray}

It should be stressed that taking the orientation vector in the opposite direction
\begin{eqnarray}
{\bf n} = (0,0, -1)\; , \qquad \Psi =  e^{-iE t / \hbar  } \;
(u_{0} - i u_{3})^{\alpha} = e^{-iE t/\hbar} \;  \cos ^{\alpha}
\rho \; e^{-i \alpha z} \;  , \label{3.7a}
\end{eqnarray}

\noindent we arrive at
\begin{eqnarray}
\Psi^{\alpha_{\pm}}_{-} = e^{-iE t/\hbar} \;  \cos^{\alpha_{\pm}}
\rho  \; e^{ -i \alpha z} \; , \qquad
 \alpha_{\pm}  =
-1 \pm \sqrt{2\epsilon+1}\; , \qquad \sqrt{2\epsilon+1}= 0, 1, 2,
... ,
\nonumber
\end{eqnarray}

\noindent from which physical ones are
\begin{eqnarray}
\alpha _{+} \in  + 1, + 2, ... \qquad \Psi^{\alpha_{+}} _{-}(\rho,
\phi, z) \; . \label{3.7c}
\end{eqnarray}

\vspace{5mm} Let us show that the plane wave solution may by considered as
an eigenfunction  in the problem
\begin{eqnarray}
({\bf P} \; {\bf n} ) \; \Psi = \alpha \; \Psi \; , \qquad \Psi =
e^{-i\epsilon t} \; (u_{0} + i{\bf n} \; {\bf u})^{\alpha} \; ,
\label{3.8a}
\end{eqnarray}
\begin{eqnarray}
{\bf P} = -i (1 + {\bf q} \bullet {\bf q}) \; {\partial \over
 \partial {\bf q}} \;, \qquad
{\bf q} = {   {\bf u} \over u_{0} } \;, \qquad u_{0}^{2} + {\bf
u}^{2} =1 \; .
\label{3.8a}
\end{eqnarray}

\noindent It should be noted that the  variable $q_{i}$ parameterizes correctly only
elliptic model, because  it does not distinguish between  $(+u_{0},+{\bf u}) $
and  $(-u_{0},-{\bf u}) $ ; for simplicity  below we perform calculation for this case.
In variable $q_{i}$:
\begin{eqnarray}
u_{0} = {1 \over + \sqrt{1 + q^{2}} }  \; , \qquad {\bf u} = {
{\bf q} \over  + \sqrt{1 +q^{2}}} \; , \qquad
\Psi =  e^{-i\epsilon t} \; (1 + q^{2} )^{-\alpha /2}   \; (1 +
i{\bf n} \; {\bf q})^{\alpha} \; , \label{3.8b}
\end{eqnarray}

\noindent for
\begin{eqnarray}
({\bf P} \; {\bf n} ) \; \Psi  =-i ( n_{i} { \partial \over
\partial q_{i}}   -
 n_{j}q_{j} \;q_{i} {\partial \over \partial q_{i}} ) \;\left [
  (1 - q^{2} )^{-\alpha /2}   \; (1 + {\bf n} \; {\bf q})^{\alpha} \right ]  =
\nonumber
\\
=  -i \; (1 + q^{2} )^{-\alpha /2}   \; (1 + i {\bf n} \; {\bf
q})^{\alpha} \; [\; -\alpha \; { (n_{i} q_{i})  \over  1 + q^{2} }
\;  + {i \alpha   \over 1 +i {\bf n} \; {\bf q} } -
 {(n_{j} q_{j})\;
\alpha q^{2} \over  1 + q^{2} } \;  +i { \alpha  \; (n_{j} q_{j})
\;
 (q_{i}n_{i})     \over 1 +  i {\bf n} \; {\bf q} } \; ]\; .
\nonumber
\end{eqnarray}

\noindent after simple calculation we get
\begin{eqnarray}
( {\bf P} \; {\bf n} )  \Psi = \alpha \; \Psi \;  . \label{3.8c}
\end{eqnarray}

\noindent Substituting  the function $\Psi$ into Schr\"{o}dinger eqaution, we relate $\alpha$ with energy $\epsilon$:
\begin{eqnarray}
\alpha = \alpha _{\pm}=  -1 \pm  \sqrt{2\epsilon +1} \; .
 \label{3.8e}
\end{eqnarray}

\noindent In usual  units one can easily perform the limiting procedure
\begin{eqnarray}
\alpha \; {\hbar \over \rho} = - {\hbar \over \rho}  \pm   {\hbar
\over \rho} \; \sqrt{ {2E   M \rho^{2}\over \hbar^{2} }
 +1}  =  -  {\hbar \over \rho}  \pm  \sqrt{2EM + {\hbar^{2} \over \rho^{2}} } \;\;
 \longrightarrow \pm \sqrt{2EM} \; \;\; \mbox{при} \;\;  \rho \rightarrow \infty \; .
\nonumber
\end{eqnarray}

Now we want to relate the plane waves on the sphere $S_{3}$ with solutions
arising in the frames of general method of separation of variables.
Let start with Schr\"{o}dinger equation for cylindric waves
\begin{eqnarray}
\Phi (\rho ,\phi ,z) = e^{+i m \phi } \; e^{+i\alpha z} \; R(\rho
) \; , \nonumber
\\
\; [ \; {d^{2} \over d\rho ^{2} } +  ( {\cos \rho \over  \sin
\rho}   - {\sin  \rho \over  \cos  \rho }) {d \over d\rho }  + 2
\epsilon  - {m^{2} \over  \sin ^{2} \rho } - { K^{2} \over \cos
^{2} \rho } \;  ] \; R(\rho ) = 0 \; . \nonumber
\end{eqnarray}

\noindent The functions  $R(\rho )$  are expressed in hypergeometric functions (for more detail see
\cite{Red'kov-2007}:
\begin{eqnarray}
R(\rho ) = \sin ^{a} \rho  \cos ^{b} \rho \;  F ( A , B , C ; \cos
^{2}\rho  )\; , \nonumber
\\
a = \pm m \; , \qquad b = \pm \alpha \; , \qquad C =  b + 1  \; ,
\nonumber
\\
A = ( a + b + 1 - \sqrt{2 \epsilon  + 1 } ) / 2  \; , \qquad
B = ( a + b + 1 + \sqrt{2 \epsilon  + 1} ) / 2 \;  . \label{3.9a}
\end{eqnarray}

\noindent To have wave functions $\Phi (\rho ,\phi ,z)$, single-valued and finite on $S_{3}$,  we must require
\begin{eqnarray}
M  = \{ \; 0, \pm 1, \pm 2, \ldots \; \} \; ,  \qquad N  = \{ \;
0, \pm 1, \pm 2, \ldots   \; \} \; , \;\; \qquad a  =  + \mid M
\mid \; , \;\; b  =  \mid K \mid \;  . \label{3.9b}
\end{eqnarray}

\noindent and make the hypergeometric series a polynomials:
\begin{eqnarray}
A  =  - n  , \;\;  n = 0, 1, 2, 3,\ldots  \; , \qquad
\epsilon _{N} = {1 \over 2} \; (N^{2}- 1)\; , \;\; N = a + b + 1 +
2 n \; ; \label{3.9c}
\end{eqnarray}

When  $\rho = 0 $ we have
\begin{eqnarray}
n_{0} = \cos  z \; , \;\; n_{1} = 0 \; , \;\; n_{2} = 0 \; , \;\;
n_{3} = \sin  z\; ; \label{3.12a}
\end{eqnarray}

\noindent   so the construct  $( 0 , \phi  , z )$  represents points
  $n_{a} = ( \cos z, 0 , 0 , \sin z )$; correspondingly, the  wave function
  $\Phi _{\epsilon m \alpha }$  behaves as follows
  \begin{eqnarray}
m \neq  0 \; , \qquad  \Phi _{\epsilon m \alpha } = 0 \;\;  ,
\nonumber
\\
m = 0 \;, \; \Phi _{\epsilon 0 \alpha } = e^{+i \alpha z} \; F(A,
B, C; 1)  = \Phi (n) \; ; \label{3.12b}
\end{eqnarray}

\noindent being a single valued and continuous on  $S_{3}$.
In the same manner, when $\rho = \pi /2$ we have
\begin{eqnarray}
n_{0}  = 0 \; , \; n_{1}  = \cos  \phi \; , \; n_{2} = \sin  \phi
\; , \; n_{3}  = 0 \; ;
\nonumber
\\
\alpha  \neq  0 \; , \qquad \qquad   \Phi _{\epsilon m \alpha }  =
0 \;\; , \nonumber
\\
K = 0 \;, \; \Phi _{\epsilon m 0} = e^{+im \phi } \; F(A, B, C; 0)
\; . \label{3.13b}
\end{eqnarray}

To obtain the Shapiro plane waves, one  must separate in
(\ref{3.9a}) a subset with  $\underline{m = 0}$,
and then restrict oneself by
$\underline{A=0}$ (what corresponds to  $n=0$),
as result we arrive at
\begin{eqnarray}
 \Phi (\rho ,\phi ,z) =
\; e^{+i\alpha z} \; \cos ^{\mid \alpha \mid} \rho \;  F ( A , B ,
C ; \cos ^{2}\rho  ) = e^{+i\alpha z} \; \cos ^{\mid \alpha \mid}
\rho  \; , \nonumber
\\
 b = \mid \alpha \mid , \;\; \alpha = 0, \pm 1, \pm 2, ... \; , \qquad C =  b + 1  \; ,
 \nonumber
\\
B = {  b + 1 + \sqrt{2 \epsilon  + 1} \over 2} \; , \qquad
A =  0  \; , \qquad
\epsilon  = {(b+1)^{2} - 1 \over 2} \; . \label{3.14}
\end{eqnarray}

The previous conclusion may be repeated:
the  general  method of separation of variables   embraces the  all plane wave solutions;
the plane waves in spherical  space  consist of  a small part
of the whole set of basis wave functions of Schr\"{o}dinger equation.

We gave  special attention to plane wave solutions in spherical model $S_{3}$, expecting  in the future
 to extend results of \cite{Bogush-Kurochkin-Otchik-2003} on  the scattering problem on Coulomb center in Lobachevsky space   to spherical model.

\subsection*{4. Plane waves in orispherical coordinates of Lobachevsky space}

As was shown in \cite{Bogush-Kurochkin-Otchik-Bychkovskaya-2006}, for Shapiro waves
the horisperical coordinate play a a special role. In \cite{Olevsky-1950} this coordinate system
has been listed under the number
 XIV\footnote{To avoid misunderstanding it should be noted that in \cite{Vilenkin-Smorodinsky-1964},
 \cite{Bogush-Kurochkin-Otchik-Bychkovskaya-2006}
instead of  $(r,\phi,z)$ were used variables $(r, \pi/2- \phi , -a,)$}:
\begin{eqnarray}
u_{1} = r e^{-z} \cos \phi \; , \qquad  u_{2} = r e^{-z} \sin \phi
\; , \nonumber
\\
u_{3} = \mbox{sh}\;z + {1 \over 2} \;  r^{2} e^{-z}   = {1 \over
2}  \; [\;  e^{+z} +( r^{2}-1) e^{-z} ] \; , \nonumber
\\
u_{0} = \mbox{ch}\;z + {1 \over 2}\; r^{2} e^{-z}   = {1 \over 2}
\; [ \; e^{+z} +(r^{2}+1) e^{-z} ] \; ; \nonumber
\\
dS^{2} = dt^{2} - e^{-2z} \; d r^{2} -  e^{-2z} r^{2}\;  d\phi^{2}
- dz^{2} \; . \label{4.1}
\end{eqnarray}

\noindent In the limit os flat space they reduce to ordinary cylindric coordinates.

Arbitrary Shapiro's  wave is determined by
\begin{eqnarray}
u_{0}^{2} - {\bf u}^{2} =1 \; , \qquad \Psi =  e^{-i\epsilon t} \;
(u_{0} + {\bf n} \; {\bf u})^{\alpha}  \; , \nonumber
\end{eqnarray}

\noindent
 taking the   orientation vector  ${\bf n}$  according to ${\bf n} = (0,0, -1)$ we get
 \begin{eqnarray}
\Psi =  e^{-iE t/\hbar } \; (u_{0} - u_{3})^{\alpha} =    e^{-iE
t/\hbar} e^{- \alpha  z} \; . \label{4.2}
\end{eqnarray}

\noindent It may be recognize as a solution of Schr\"{o}dinger equation in $H_{3}$
\begin{eqnarray}
  i    \;  \partial  _{t }     \Psi =
   - { \hbar^{2}\over 2 M\rho^{2}  } \left [ \; { e^{2z} \over r} {\partial \over r} r {\partial \over \partial r} +
 {e^{2z} \over
r^{2} } {\partial^{2} \over \partial \phi^{2} } + {1 \over e^{-2z}
} {\partial \over \partial z}  e^{-2z} {\partial \over \partial z}
\; \right ] \; \Psi \; ; \nonumber
\end{eqnarray}

\noindent which with the substitution \ref{4.2})  gives
\begin{eqnarray}
2\epsilon e^{-\alpha z} = ({\partial^{2} \over \partial z^{2}} -
2{\partial \over \partial z}) e^{-\alpha z} \; ,
\qquad
\alpha^{2} - 2\alpha + 2\epsilon = 0 \; ,
 \nonumber
\\
  \alpha = 1 \; \mp \; i \; \sqrt{2\epsilon -1}\; ;\; , \qquad
\Psi =   e^{-iE t/\hbar} e^{(-1 \; \pm\;  i \; \sqrt{2\epsilon -1}
)z} \; . \label{4.3}
\end{eqnarray}

Let us find exptression for momentum operator  $P_{3}$ in these coordinates
 $x^{i} =(r, \phi, z)$. Starting with
\begin{eqnarray}
q_{1} = {2r \;  \cos \phi \over e^{2z} +r^{2}+1  }\; , \qquad
q_{2} = {2r \;  \sin \phi \over e^{2z} +r^{2}+1  }\; ,\qquad q_{3}
= { e^{2z} +r^{2} -1 \over e^{2z} +r^{2}+1 }\; ,` \label{4.4a}
\end{eqnarray}

\noindent  and inverse ones
\begin{eqnarray}
r =  { \sqrt{q_{1}^{2} +q_{2}^{2}}  \over 1 -q_{3}} \; ,  \qquad
e^{2z} = {1 -q^{2} \over (1-q_{3})^{2}} \; , \nonumber
\\
\cos \phi = { q_{1} \over  \sqrt{q_{1}^{2} +q_{2}^{2}}}\; , \qquad
\sin \phi = { q_{2} \over  \sqrt{q_{1}^{2} +q_{2}^{2}}}\; .
\label{4.4c}
\end{eqnarray}

Further, with the help of formulas
\begin{eqnarray}
{\partial r \over \partial q_{1} } = {1 \over 2}( e^{2z} +r^{2}
+1) \cos \phi \; , \; {\partial r \over \partial q_{2} } = {1
\over 2}( e^{2z} +r^{2} +1) \sin \phi \; , \; {\partial r \over
\partial q_{3} } = {1 \over 2}\; r\; ( e^{2z} +r^{2} +1) \; ,
\nonumber
\\
{\partial \phi \over \partial q_{1}} = -{(e^{2z} +r^{2}+1 )\over
2r} \; \sin \phi \; , \qquad {\partial \phi \over \partial q_{2}}
= {(e^{2z} +r^{2}+1 )\over 2r} \; \cos \phi \; , \qquad {\partial
\phi \over \partial q_{3}}  = 0 \; , \nonumber
\\
{\partial z \over \partial q_{1}} = - {1 \over 2} \;     e^{-2z} r
(e^{2z} +r^{2}+1) \cos \phi \; , \qquad {\partial z \over \partial
q_{2}} =  - {1 \over 2} \;     e^{-2z} r  (e^{2z} +r^{2}+1) \sin
\phi  \; , \nonumber
\\
{\partial z \over \partial q_{3}} =  {1\over 4}e^{-2z} [(e^{2z}
+1)^{2} -r^{4}]  \; .
\nonumber
\end{eqnarray}

\noindent after simple calculation  we get
\begin{eqnarray}
 P_{3} =-i \; [
   {\partial  x^{i} \over  \partial q_{3}} {\partial \over \partial x^{i}} -
  q_{3}\;
( q_{1}  {\partial x^{i} \over \partial q_{1}}  + q_{2}  {\partial
x^{i} \over \partial q_{2}}  + q_{3}  {\partial x^{i} \over
\partial q_{3}}  ) {\partial \over  \partial x^{i}}  ] =
  -i ( \; r \; {\partial \over \partial r} +  {\partial
\over  \partial z } \; )  \; . \label{4.9}
\end{eqnarray}

Therefore, the plane wave is an eigenfunction of $P_{3}$:
\begin{eqnarray}
{\bf n} = (0,0, -1)\; , \qquad \Psi_{-}  =  e^{-iE t/\hbar } \;
(u_{0} - u_{3})^{\alpha} =
\nonumber
\\
=  e^{-iE t/\hbar} e^{- \alpha  z} \;  , \qquad P_{3} \Psi_{-} = +i \alpha \;  \Psi_{-}\; .
\label{4.10a}
\end{eqnarray}

\noindent
It should be  noted a plane wave with opposite direction:
\begin{eqnarray}
{\bf n} = (0,0, +1)\; , \qquad \Psi_{+} =  e^{-iE t/\hbar } \; (
u_{0} + u_{3})^{\alpha} =
\nonumber
\\
=  e^{-iE t/\hbar} ( \; e^{z} + r^{2}
e^{-z} \; ) ^{\alpha  } \; ,
\qquad P_{3} \Psi_{+} = - i \alpha \; \Psi_{+} \; ;
\label{4.10b}
\end{eqnarray}

\noindent
Therefore,  both $\Psi_{\pm}$  should be considered as representing  plane waves.
Also, we may verify by simple calculation that $\Psi_{+}$ satisfies the Schr\"{o}dinger equation as  well.
\begin{eqnarray}
  i    \;  \partial  _{t }     \Psi _{+}=
   - { \hbar^{2}\over 2 M\rho^{2}  }  [ \; { e^{2z} \over r} {\partial \over r} r {\partial \over \partial r} +
 {e^{2z} \over
r^{2} } {\partial^{2} \over \partial \phi^{2} } + {1 \over e^{-2z}
} {\partial \over \partial z}  e^{-2z} {\partial \over \partial z}
\;  ] \;\Psi _{+} \; ; \nonumber
\\
\Psi_{+} =     e^{-iE t/\hbar} ( \; e^{z} + r^{2} e^{-z} \; )
^{\alpha  } \; , \qquad \alpha = 1 \; \mp \; i \; \sqrt{2\epsilon
-1}\; . \nonumber
\end{eqnarray}

\subsection*{5.   Complex orispherical coordinates in $S_{3}$ }

To introduce  complex horisperical coordinates is spherical space  $S_{3}$, let us start with
real  horisperical coordinates (\ref{3.1})  is Lobachevsky  space  $H_{3}$
\begin{eqnarray}
u_{1} = r e^{-z} \cos \phi \; , \qquad  u_{2} = r e^{-z} \sin \phi
\; , \nonumber
\\
u_{3} =  {1 \over 2}  \; [\;  e^{+z} +( r^{2}-1) e^{-z} ] \; ,
\qquad u_{0} =  {1 \over 2} \; [ \; e^{+z} +(r^{2}+1) e^{-z} ] \;
. \label{5.1}
\end{eqnarray}

\noindent
Inverse formulas are
\begin{eqnarray}
\cos \phi = {u_{1} \over \sqrt{u_{1}^{2} + u_{2}^{2} } } \;
,\qquad \sin \phi = {u_{2} \over \sqrt{u_{1}^{2} + u_{2}^{2}}} \;
, \nonumber
\\
e^{z} = {1 \over u_{0} - u_{3} } \; , \qquad r^{2} = { u_{0} +
u_{3} \over u_{0} - u_{3} } -  {1 \over ( u_{0} - u_{3} )^{2} } \;
. \qquad r^{2}  + e^{2z} = { u_{0} + u_{3} \over u_{0} - u_{3} }
\; . \label{5.2}
\end{eqnarray}

\noindent Transition to spherical model can be realized for formal change:
\begin{eqnarray}
u_{0} = V_{0} \; , \qquad {\bf u} = i \; {\bf V} \; , \qquad
V_{0}^{2} + {\bf V}^{2} = 1 \; ; \nonumber
\end{eqnarray}

\noindent therefore, relations defining complex orispherical coordinates in $S_{3}$ are
\begin{eqnarray}
\cos \phi = {V_{1} \over \sqrt{V_{1}^{2} + V_{2}^{2} } } \;
,\qquad \sin \phi = {V_{2} \over \sqrt{V_{1}^{2} + V_{2}^{2}}} \;
, \nonumber
\\
e^{z} = {1 \over V_{0} - iV_{3} } \; , \qquad r^{2} = { V_{0} + i
V_{3} \over V_{0} - iV _{3} } -  {1 \over ( V_{0} - iV_{3} )^{2} }
\; . \label{5.3}
\end{eqnarray}

To parameterize real space $S_{3}$, one  must impose restrictions on
complex coordinate. Evidently,  $\phi$  is real-valued coordinate; besides, allowing for identitie
$V_{0} -iV_{3} = e^{-z} \;, \; V_{0} +iV_{3} = e^{-z^{*}} \;,
$
 we arrive at the following relationship for $r,z$:
\begin{eqnarray}
r^{2} = e^{z-z^{*}} - e^{2z} \; . \label{5.4}
\end{eqnarray}

\noindent Let us introduce notation for complex variables:
\begin{eqnarray}
r =A + iB \;, \qquad z = a + ib \; , \nonumber
\end{eqnarray}

\noindent Eqs.  (\ref{5.4}) take the real form
\begin{eqnarray}
A^{2} -B^{2} = - \cos 2b \; (e^{2a}-1) \; , \qquad
  2AB = -\sin 2b \; (e^{2a}-1)\; .
\label{5.5}
\end{eqnarray}

\noindent from whence we get \footnote{Second solution  $ A^{2}
+B^{2} = -(e^{2a} -1) $ must be rejected.}:
\begin{eqnarray}
+(e^{2a} -1  ) = A^{2} +B^{2}\; , \qquad  a \in [ \;  0, + \infty
) \; , \nonumber
\\
 \cos 2b = - {A^{2} -B^{2} \over  A^{2} + B^{2} } \; , \qquad
  \sin 2b =  - {2AB   \over  A^{2}  + B^{2} }  \; .
\label{5.6a}
\end{eqnarray}

\noindent
In turn, eqs. (\ref{5.6a}) give
\begin{eqnarray}
r^{2} = (A+iB)^{2}= - (e^{a} -1) (\cos 2b + i \sin 2b) = -rr^{*}
e^{z-z^{*}} \; ,
\nonumber
\end{eqnarray}

\noindent therefore eq.  (\ref{5.4})  may be transformed into
\begin{eqnarray}
 e^{2z} = - {r  \over r^{*}}  -r^{2} \;, \;\; \Longrightarrow \;\;  (e^{2z} +  r^{2})  (e^{2z} +  r^{2})^{*} = 1\;.
\label{5.7a}
\end{eqnarray}

Eqs.  (\ref{5.5}) can be easily solved with respect to variables  $A,B$\footnote{In general,  $(A,B)$
can be found up to $\pm$ sign.}:
\begin{eqnarray}
A =  -\sqrt{e^{2a} -1 }\; \sin b \; , \qquad B = +
\sqrt{e^{2a}-1}\; \cos b \; . \label{5.8a}
\end{eqnarray}

\noindent Below, we will use $(a,b)$ as two real independent parameters:
\begin{eqnarray}
z = a+ i b \; , \quad r = i\; \sqrt{e^{2a} -1}\;  e^{i b} \; ю
\label{5.9}
\end{eqnarray}

\noindent Inverse to (\ref{5.3})  are (compare with  (\ref{5.1})):
\begin{eqnarray}
iV_{1} = r e^{-z} \cos \phi \; , \qquad  \qquad  iV_{2} = r e^{-z} \sin
\phi \; , \nonumber
\\
iV_{3} =  {1 \over 2}  \; [\;  e^{+z} +( r^{2}-1) e^{-z} ] \; ,
\qquad
V_{0} =  {1 \over 2} \; [ \; e^{+z} +(r^{2}+1) e^{-z} ] \; .
\label{5.10}
\end{eqnarray}

The rule to perform a limiting procedure in  (\ref{5.10}) to the flat space model
is given by
\begin{eqnarray}
\rho V_{1} \longrightarrow x\; , \qquad \rho V_{2} \longrightarrow
y\; , \qquad \rho V_{3} \longrightarrow z\; , \nonumber
\\
(-i\rho r) \longrightarrow r \; , \qquad \phi \longrightarrow \phi
\; , \qquad (-i z \rho ) \longrightarrow z \nonumber
\\
x = r \cos \phi \; , \qquad y = r \sin \phi \; , \qquad z = z  \;
. \label{5.11c}
\end{eqnarray}

\noindent From  (\ref{5.10}) it follows
\begin{eqnarray}
 V_{1} =  \sqrt{e^{2a}-1}\;  e^{-a} \cos \phi \; , \qquad
 V_{2} =    \sqrt{e^{2a}-1}\;  e^{-a} \sin  \phi \; ,
\nonumber
\\
V_{3} =  e^{-a} \; \sin b \; , \qquad V_{0} =   e^{-a} \; \cos b
\; . \label{5.11a}
\end{eqnarray}

\noindent in turn, eqs. (\ref{5.11a}) give
\begin{eqnarray}
V_{1}^{2} + V_{2}^{2} = 1 -e^{-2a} \; , \qquad {V_{2} \over V_{1}}
= \mbox{tg}\; \phi \; , \qquad V_{0}^{2} + V_{3}^{2} = \; e^{-2a}
\; , \; \qquad {V_{3} \over V_{9}} = \mbox{tg}\; b \; .
\label{5.11b}
\end{eqnarray}

We may note two peculiarities in coordinates $(a , b,
\phi )$:
\begin{eqnarray}
a \in [\; 0 , \; + \infty \; )\; , \qquad b \in [\; 0, 2\pi \; ]\;
, \qquad  \phi \in [ \; 0 , 2 \pi \; ] \; , \nonumber
\\[3mm]
a \rightarrow 0\; , \qquad V_{1} = 0 \; , \; V_{2}=0\;,  \qquad
V_{0}^{2} + V_{3}^{2} = 1 \; ; \nonumber
\\
V_{0} = \cos b \; , \qquad V_{3} = \sin b\; , \qquad \phi
-\mbox{mute variable} \nonumber
\\
a \rightarrow + \infty \; , \qquad V_{0} = 0 \; , \; V_{3}=0\;,
\qquad  V_{1}^{2} + V_{2}^{2} = 1 \; ; \nonumber
\\
V_{1} = \cos \phi  \; , \qquad V_{2} = \sin \phi \; , \qquad b
-\mbox{mute variable } \; . \label{5.11e}
\end{eqnarray}

The identity  (\ref{5.4}) permits us to express  all four real coordinates
 $(V_{0},{\bf V})$ given by  (\ref{5.10}) in terms of variables  $(z,z^{*},\phi)$:
\begin{eqnarray}
V_{1} =   \sqrt{ 1 - e^{-z-z^{*} }  }  \;  \cos \phi \; , \qquad
V_{2} =\sqrt{ 1 - e^{-z-z^{*} } }  \;  \sin \phi \; , \nonumber
\\
V_{3} =    { e^{ -z^{*} }  - e^{-z} \over 2  i} \; , \qquad V_{0}
=   { e^{ -z^{*} }  + e^{-z}  \over 2  }  \; . \label{5.12b}
\end{eqnarray}

Alternatively, with the help of identities
 (see  (\ref{5.7a}))
\begin{eqnarray}
e^{z} = i \sqrt{  \; (1 + r r^{*}) \; {r \over r^{*}} \; } \; ,
\qquad e^{-z} = -i \sqrt{  \; { 1 \over 1 + r r^{*} }  \; {r^{*}
\over r  }  \; } \; , \nonumber
\end{eqnarray}

\noindent all four coordinates  $(V_{0}, {\bf V})$ (\ref{5.10}) can be
expressed in terms of variables
 $(r,r^{*},\phi)$:
 \begin{eqnarray}
V_{1} =     -  \; \sqrt{   { rr^{*} \over 1 + r r^{*} }  } \cos
\phi \; , \qquad  \qquad
 V_{2} =  -  \; \sqrt{   { r r^{*} \over 1 + r
r^{*} }  }   \sin \phi \; ,
\nonumber
\\
V_{3} =
   { r  +  r^{*} \over 2}  \sqrt{  \; { 1 \over 1 + r r^{*} }  \; {1 \over r r^{*}  }  \; } \; , \qquad
   V_{0} =   { r^{*} - r \over 2i}  \sqrt{  \; { 1 \over 1 + r r^{*} }  \; {1 \over r   r^{*}}  \; }
 \; .
\label{5.12c}
\end{eqnarray}

\subsection*{6.  Plane waves in complex  coordinates of $S_{3}$ }

Let us specify Shapiro plane waves (for shortness we will omit the factor   $e^{-i E t / \hbar } $):
\begin{eqnarray}
V_{0}^{2} + {\bf V}^{2} =1 \; , \qquad \Psi =   (V_{0} + i\; {\bf
n} \; {\bf V})^{\alpha}  \; . \nonumber
\end{eqnarray}

\noindent in complex orispherical coordinates.
Taking two opposite orientations we get
\begin{eqnarray}
{\bf n} = (0, 0, -1)\; , \qquad \Psi_{-} =  (V_{0} - i\;
V_{3})^{\alpha}  =
  e^{-\alpha z} =  e^{-\alpha (a+ i  b)} \; ,
\nonumber
\\[3mm]
{\bf n} = (0, 0, +1)\; , \qquad \Psi_{+} =   (V_{0} + i\;
V_{3})^{\alpha}  =
 e^{-\alpha z^{*}}=
    (e^{z}  + r^{2} e^{-z})^{\alpha} = e^{-\alpha (a-ib)} \; .
\label{6.13}
\end{eqnarray}

\noindent These solutions are eigenfunctions of   $P_{3}$ in  $S_{3}$:
\begin{eqnarray}
P_{3} = -i \; ( \; r \;{\partial \over r} + {\partial \over
\partial z} \; ) \; , \qquad
P_{3} \Psi_{-} = +i\alpha \; \Psi_{-} \; , \qquad P_{3} \Psi_{+} =
-i\alpha \; \Psi_{+} \; . \label{6.14}
\end{eqnarray}

\noindent we should note the expression for  $\psi_{+}$ in
(\ref{6.13}) as a function of conjugate  variable $z^{*}$, which agrees with
the use of variables
$(z,z^{*},\phi)$   or $(r,r^{*},\phi)$ as independent ones. We might expect other symmetric variant
in variables  $r,r^{*}$; indeed,
\begin{eqnarray}
e^{-z} = \sqrt{ -{r^{*} \over r} \; {1 \over 1 + rr^{*}}} \; ,
\qquad \Longrightarrow \qquad \Psi _{-} =  \left [ \; \sqrt{
-{r^{*} \over r} \; {1 \over 1 + rr^{*}}} \; \right ]^{\alpha} \;
; \label{6.15a}
\end{eqnarray}
\begin{eqnarray}
e^{z} +r^{2} e^{-z} =  \sqrt{ -{r^{*} \over r} \; {1 \over 1 + rr^{*}}} \; \;\; (-{r
\over r^{*}}) = \sqrt{ -{r \over r^{*}} \; {1 \over 1 + rr^{*}}}
\; , \;\; \Longrightarrow
\nonumber
\\
\Psi_{+} =  \left [\; \sqrt{ -{r \over r^{*}} \; {1 \over 1 +
rr^{*}}}  \; \right ] ^{\alpha} \; . \label{6.15b}
\end{eqnarray}

Now, let us turn  to Schr\"{o}dinger equation in complex coordinates.
We are to get the metric of space in these coordinates, staring from
\begin{eqnarray}
dS^{2}  = dt^{2} - dl^{2}\; , \qquad dl^{2} = dV_{0}^{2} +
dV_{1}^{2} + dV_{2}^{2} + dV_{3}^{2} \nonumber
\end{eqnarray}

\noindent and allowing for identities
\begin{eqnarray}
 d V_{1} = -i\;  d \;(   r e^{-z} \cos \phi ) = -i \; e^{-z} [\;
   \cos \phi \;(  dr  -  r  \; dz  ) - r  \sin  \phi  \; d \phi \;  ) \; ,
\nonumber
\\
 d V_{2} = -i\;  d \;(    r e^{-z} \sin \phi ) = -i \; e^{-z} [\;
   \sin \phi \; ( dr  -  r \; dz )  + r  \cos  \phi  \; d \phi  \; ) \; ,
\nonumber
\\
dV_{3} =  -i \; {1 \over 2}  \;
 [\; ( e^{+z}\;  dz +         e^{-z}  2  R\; dr  - r^{2} e^{-z} \; dz ) +   e^{-z} dz   \;  ] \; ,
\nonumber
\\
dV_{0} =   {1 \over 2}  \;   [\; ( e^{+z}\;  dz +         e^{-z}
2  r\; dr - r^{2} e^{-z} \; dz)  -  e^{-z} dz   \;  ] \; ,
 \nonumber
\end{eqnarray}

\noindent we get
\begin{eqnarray}
dS^{2} = dt^{2} + e^{-2z} \; d r^{2} +  e^{-2z} r^{2}\;  d\phi^{2}
+ dz^{2} \; , \qquad \sqrt{-g} =  \sqrt{-e^{4z} \; r^{2}} = i r\;
e^{-2z} \; ; \label{6.16b}
\end{eqnarray}

\noindent take notice on four signs $++++$ in metrical tensor. Correspondingly,
Schr\"{o}dinger Hamiltonian is given by
\begin{eqnarray}
H = { \hbar^{2} \over 2 M^{2} \rho^{2} } \; {1 \over \sqrt{-g} }
\partial_{i}  \sqrt{-g} g^{ij}
\partial_{j} =
 + { \hbar^{2}\over 2 M\rho^{2}  }  [ \; { e^{2z} \over r} {\partial \over r} r {\partial \over \partial r} +
 {e^{2z} \over r^{2} } {\partial^{2} \over \partial \phi^{2} } + {1
\over e^{-2z} } {\partial \over \partial z}  e^{-2z} {\partial
\over \partial z} \;  ] \; . \label{6.17}
\end{eqnarray}

One may easily verify that above constructed plane waves satisfy the Sschr\"{o}dinger equation (\ref{6.17}).
It is so for the wave  $\Psi_{\pm}$:
\begin{eqnarray}
\Psi_{-} = e^{-Et / \hbar}
  e^{-\alpha z} \;, \qquad 2\epsilon  e^{-\alpha z}  =  ({d^{2} \over dz^{2}} - 2{d \over dz} ) e^{-\alpha z} \; ,
\nonumber
\\
\alpha^{2} +2 \alpha -2\epsilon = 0 \; , \qquad \alpha = -1 \pm
\sqrt{2\epsilon +1}\; . \label{6.18}
\end{eqnarray}

\noindent For the  eave with different orientation
we have an equation
\begin{eqnarray}
2\epsilon (e^{z}  + r^{2} e^{-z})^{\alpha} = [ \;  e^{2z}
{\partial^{2} \over \partial r^{2} }  +  {e^{2z} \over r}
{\partial \over \partial r} + {\partial ^{2} \over \partial z^{2}}
-2 {\partial \over \partial z} \; ] (e^{z}  + r^{2}
e^{-z})^{\alpha} \; .
\label{6.20a}
\end{eqnarray}

\noindent which is identity if
\begin{eqnarray}
\alpha^{2} + 2\alpha = 2\epsilon \; , \qquad \alpha = -1 \pm
\sqrt{2\epsilon +1}\; .
\nonumber
\end{eqnarray}

The plane wave (\ref{6.20a}), as well as and its counterpart  in  $H_{3}$ , can hardly be considered as constructed in the
frame of general method of separation of variables.

Now we should  examine the plane waves constructed with respect  to their continuity  properties
in the  space $S_{3}$:
\begin{eqnarray}
\Psi_{-} =   e^{-\alpha z}  = e^{-\alpha (a+ib)} \;,  \qquad
\Psi_{+} =   e^{-\alpha z^{*}}  = e^{-\alpha (a-ib)}\;,
\qquad
   \alpha = -1 \pm \sqrt{2\epsilon +1}\; ;
\nonumber
\\
 V_{1} =  \sqrt{e^{2a}-1}\;  e^{-a} \cos \phi \; , \;
 V_{2} =    \sqrt{e^{2a}-1}\;  e^{-a} \sin  \phi \; ,
\; V_{3} =  e^{-a} \; \sin b \; , \; V_{0} =   e^{-a} \; \cos b
\; .
\nonumber
\end{eqnarray}

\noindent
Evidently, one must require $\Psi_{\pm}$ to be $2\pi$-periodic  in variable  $b$:
\begin{eqnarray}
\alpha = -1 +  \sqrt{2\epsilon +1} =  n = 0, +1, +2, ... : \qquad
\epsilon ={(n+1)^{2} -1 \over 2} \; ; \nonumber
\\
\alpha = -1 -  \sqrt{2\epsilon +1} =  n = -2, -3,... : \qquad
\epsilon ={(n+1)^{2} -1 \over 2} \; ; \label{6.21}
\end{eqnarray}

\noindent There arise four types of solutions:
\begin{eqnarray}
\Psi_{-}^{\alpha \geq 0} = e^{(1-\sqrt{2\epsilon +1} )\; a} \;\;
e^{ +\;i\; (1-\sqrt{2\epsilon +1}) \; b } \; , \qquad
\Psi_{-}^{\alpha \leq 0} = e^{(1+\sqrt{2\epsilon +1} )\; a} \;\;
e^{ +\;i\; (1+\sqrt{2\epsilon +1}) \; b } \; , \nonumber
\\
\Psi_{+}^{\alpha \geq 0} = e^{(1-\sqrt{2\epsilon +1} )\; a} \;\;
e^{ -\; i\; (1-\sqrt{2\epsilon +1}) \; b } \; , \qquad
\Psi_{+}^{\alpha \leq 0} = e^{(1+\sqrt{2\epsilon +1} )\; a} \;\;
e^{ -\;i \; (1+\sqrt{2\epsilon +1}) \; b } \; .
\nonumber
\end{eqnarray}

\noindent Having remembered  he peculiarities of these  coordinates -- see (\ref{6.11e}), we must  conclude that
 second and fourth solutions  are to be rejected.  Thus, physical solutions are
 \begin{eqnarray}
\Psi_{-}^{\alpha > 0} = e^{(1-\sqrt{2\epsilon +1} )\; a} \;\;  e^{
+\;i\; (1-\sqrt{2\epsilon +1}) \; b } \; , \nonumber
\\
\Psi_{+}^{\alpha > 0} = e^{(1-\sqrt{2\epsilon +1} )\; a} \;\;  e^{
-\; i\; (1-\sqrt{2\epsilon +1}) \; b } \; ; \label{6.22c}
\end{eqnarray}

\noindent they are related by complex conjugation. In a particular case,
 $\epsilon =0, \; \alpha = 0$, we have very specific solution
\begin{eqnarray}
\epsilon =0, \; \alpha = 0 \; , \qquad \Psi^{\epsilon = 0} _{\pm}
=(V_{0} \pm iV_{3})^{0} = 1 \; . \label{6.22e}
\end{eqnarray}

\noindent which represents  a quantum state with uniform probability distribution in
spherical space  $S_{3}$.

Having in mind the possibility to express the plane waves in variable $z,z^{*}$  or
$(r,r^{*}$), or $(a,b)$ let us consider the task of translating the metric tensor to those
coordinates.

First, with the help of relation
$r^{2} = e^{z-z^{*}} - e^{2z} $
one  can  exclude the cariable $r$, it results in
\begin{eqnarray}
dS^{2} = dt^{2}  -     { e^{z+z^{*} } -1 \over  e^{z+z^{*} } } \;
d\phi^{2}  -   {1 \over 4 e^{z+z^{*}} (e^{z+z^{*}} - 1   ) }
 \; [ \;    dz ^{2}  +   dz^{*2}       +
 2 \; (2 e^{z+z^{*}}      -  1 )\; dz dz^{*}  \; ] \; .
 \label{6.24}
 \end{eqnarray}

\noindent Allowing for
 $z =a + ib$, we produce
 \begin{eqnarray}
dS^{2} = dt^{2}  -  { e^{2a}- 1 \over e^{2a}} \; d\phi^{2}  -
 { da^{2} \over e^{2a} - 1}\;   - {db^{2} \over e^{2a} } \; , \qquad
 \sqrt{-g}= {1 \over e^{2a}} \;  .
\label{6.25}
\end{eqnarray}

\noindent Corresponding Schr\"{o}dinger equation
is
\begin{eqnarray}
-2\epsilon \; \Psi = (\;  {e^{2a} \over e^{2a}- 1} \partial^{2}
_{\phi} +
 e^{2a}\partial_{a} {e^{2a} -1  \over e^{2a} }    \partial_{a} +e^{2a} \partial^{2}_{b} \; ) \; \Psi \; .
\label{6.26}
\end{eqnarray}

\noindent The plane waves  satisfy this equation, indeed
\begin{eqnarray}
\Psi = \Psi_{\mp} = e^{- \alpha (a \pm i\; b)} \; , \nonumber
\\
-2\epsilon \;  e^{- \alpha (a \pm i\; b)} = [\; (e^{2a} -1 )
\partial^{2}_{a}  +2 \partial_{a}  + e^{2a} \partial^{2}_{b} \; ]
\; e^{- \alpha (a \pm i\; b)} \; , \label{6.27a}
\end{eqnarray}

\noindent or
\begin{eqnarray}
-2\epsilon \; e^{- \alpha (a \pm i\; b)} = [\; (e^{2a} -1)
\alpha^{2}     -2 \; \alpha   - e^{2a} \alpha^{2} \; ] \; e^{-
\alpha (a \pm i\; b)} \; ; \nonumber
\end{eqnarray}

\noindent and further
\begin{eqnarray}
\alpha^{2} + 2\alpha -2\epsilon = 0 \; , \qquad \alpha = -1  \pm
\sqrt{2\epsilon +1} \; .
\nonumber
\end{eqnarray}

\noindent
In the same manner let us specify the  Schr\"{o}dinger equation in variables
$(z,z^{*},\phi)$. With notation
$
(z, z^{*})  = (z, W) \; $, the metyric
 (\ref{6.24}) reads
 \begin{eqnarray}
f = e^{z+W} \; , \qquad dS^{2} = dt^{2}  -     { f -1 \over  f }
\;  d\phi^{2}  - \nonumber
\\
- {1 \over 4 f (f - 1   ) }
 \; [ \;    dz ^{2}  +   d W^{2}       +
 2 \; (2 f      -  1 )\; dz d W  \; ] \; .
 \label{6.28a}
 \end{eqnarray}

\noindent
Allowing for relations
\begin{eqnarray}
g_{\alpha \beta} = \left | \begin{array}{cccc}
1 &  0  & 0  & 0 \\[2mm]
0 &  -{1 \over 4 f (f - 1)}   &   -{2f-1 \over 4f(f-1)}  & 0  \\[2mm]
0 &  - {2f-1 \over 4f(f-1)}   &  -{1 \over 4 f (f - 1)}  & 0. \\[2mm]
0 &   0  &  0  & -     { f -1 \over  f }
\end{array} \right | \; , \qquad \mbox{det} \; (g_{\alpha \beta})  = {1 \over 4f^{2}}  \; .
\nonumber
\\
g^{\alpha \beta} = \left | \begin{array}{cccc}
1 &  0  & 0  & 0 \\[2mm]
0 &   1   &   -(2f-1) & 0  \\[2mm]
0 &   -(2f-1)   &  1  & 0. \\[2mm]
0 &   0  &  0  & -     { f \over  f-1 }
\end{array} \right | \; .
\label{6.28b}
\end{eqnarray}

\noindent the Schr\"{o}dinger equation
\begin{eqnarray}
2\epsilon \Psi = [ \; {1 \over \sqrt{g} }  \partial_{\phi}
\sqrt{g} g^{\phi \phi} \partial_{\phi} + {1 \over \sqrt{g} }
\partial_{z} \sqrt{g} g^{WW} \partial_{z} + {1 \over \sqrt{g} }
\partial_{W} \sqrt{g} g^{WW} \partial_{W} + \nonumber
\\
+ {1 \over \sqrt{g} }  \partial_{z} \sqrt{g} g^{ZW} \partial_{W} +
{1 \over \sqrt{g} }  \partial_{W} \sqrt{g} g^{ZW} \partial_{z} \;
] \Psi \nonumber
\end{eqnarray}

\noindent reduces to the form
\begin{eqnarray}
2\epsilon \Psi = [ \; -  {f \over f-1}  \partial^{2}_{\phi}  + f
\partial_{z} {1 \over f}  \partial_{z} + f  \partial_{W} {1 \over
f}  \partial_{W} - \nonumber
\\
- 2f \partial_{z} {1 \over 2f} (2f-1) \partial_{W} - 2f
\partial_{W} {1 \over 2f} (2f-1) \partial_{z} \; ] \Psi \; .
\nonumber
\end{eqnarray}

\noindent Further,  taking into account identities
\begin{eqnarray}
f =e^{z+W}\;  , \qquad {1 \over f} =  e^{-z-W}\;  , \qquad
f\partial _{z} {1 \over f} = - 1  \; , \qquad f\partial _{W} {1
\over f} = - 1  \; , \nonumber
\\
-2f \partial_{z} {1 \over 2f} (2f-1)= -1 \; , \qquad -2f
\partial_{W} {1 \over 2f} (2f-1)= -1 \; . \nonumber
\end{eqnarray}

\noindent we translate  the above  equation to the form
\begin{eqnarray}
2\epsilon \Psi = [ \; -  {f \over f-1}  \partial^{2}_{\phi}  +
 \partial^{2}_{z}  - \partial_{z} +
  \partial^{2}_{W} -   \partial_{W} +
\nonumber
\\
- (2f-1) \partial_{z} \partial_{W} - \partial_{W} - (2f-1)
\partial_{W} \partial_{z} - \partial_{z} \; ]\; \Psi \; ;
\label{6.29}
\end{eqnarray}

\noindent remembering on $W = z^{*}$. Two plane wave satisfy eq. (\ref{6.29}):
\begin{eqnarray}
\Psi _{-} = e^{-\alpha z} \; , \qquad 2\epsilon \; e^{-\alpha z} =
[ \;
 \partial^{2}_{z}  - 2\partial_{z}   \; ]\; e^{-\alpha z} \; ,
\nonumber
\\
\Psi _{+} = e^{-\alpha W} \; , \qquad 2\epsilon \; e^{-\alpha W} =
[ \;
 \partial^{2}_{W}  - 2\partial_{W}  \; ]\; e^{-\alpha W} \; ,
\nonumber
\\
2\epsilon = \alpha^{2} +2\alpha \; ,  \qquad \alpha = -1 \pm
\sqrt{2\epsilon +1}\; . \label{6.30b}
\end{eqnarray}

At last, one can readily translate metrical tensor to the variables $(r,r^{*},\phi)$:
\begin{eqnarray}
dS^{2} =
 dt^{2}   + {    1      \over  4(1 + r r^{*} )^{2} } \; { d r^{2} \over  r^{2}   } +
{    1      \over  4(1 + r r^{*} )^{2} } \; { dr^{*2} \over
r^{*2}   } - \nonumber
\\
-  2 \; { (2 r r^{*}  +1 )   \over  4(1 + r r^{*} )^{2}} \;  {1\over
r  r^{*}  } \; d r \;  d r^{*} - { r r^{*} \over 1 + r r^{*}} \;
d\phi^{2} \; , \label{6.33a}
\end{eqnarray}

\noindent One could produce corresponding form of Schr\"{o}dinger Hamiltonian
and then verify that expressed in variables $r,r^{*}$ plane waveare exact solution of quantum mechanical equation
in these variables.

Let us summarize results.

The  general  method
of separation of variables   embraces the  all plane wave
solutions; the plane waves in Lobachevsky  and Riemann space
consist of  a small part of the whole set of basis wave functions
of Schr\"{o}dinger equation.

In space of constant positive curvature  $S_{3}$, a complex analog
of orispherical coordinates of Lobachevsky space $H_{3}$ is introduced.
To parameterize  real space   $S_{3}$, two complex coordinates  $(r,z)$ must obey
additional restriction in the form of  the  equation
$r^{2} = e^{z-z^{*}} - e^{2z} $.
The metrical tensor of space $S_{3}$ is expressed in terms of  $(r,z)$  with additional constraint,
or through pairs  of conjugate variables  $ (r,r^{*})$ or  $(z,z^{*})$; correspondingly exist three different representations
for Schr\"{o}dinger Hamilto\-nian.
 Shapiro plane waves are determined and explored as solutions of  Schr\"{o}dinger
 equation in complex horisperical coordinates of $S_{3}$.
 In particular, two oppositely directed plane waves may be presented
 as exponentials in conjugated  coordinates.
       $\Psi_{-}= e^{-\alpha z}$ and  $\Psi_{+}= e^{-\alpha z^{*}}$.
Solutions constructed
 are single-valued, finite, and continuous functions in spherical space and correspond to discrete energy levels.

\end{document}